# TENDENCY OF SPHERICALLY IMPLODING PLASMA LINERS FORMED BY MERGING PLASMA JETS TO EVOLVE TOWARD SPHERICAL SYMMETRY

J. T. Cassibry[1], M. Stanic[1], S. C. Hsu[2], S.I. Abarzhi[3], F. D. Witherspoon[4]

1. Propulsion Research Center, Technology Hall S-226 The University of Alabama in Huntsville, Huntsville, AL 35899, USA
2. Physics Division, Los Alamos National Laboratory, Los Alamos, NM 87545, USA
3. The University of Chicago, Chicago, IL 60637, USA
4. HyperV Technologies Corp., Chantilly, VA 20151, USA

Three dimensional hydrodynamic simulations have been performed using smoothed particle hydrodynamics (SPH) in order to study the effects of discrete jets on the processes of plasma liner formation, implosion on vacuum, and expansion. The pressure history of the inner portion of the liner was qualitatively and quantitatively similar from peak compression through the complete stagnation of the liner among simulation results from two one dimensional radiation-hydrodynamic codes, 3D SPH with a uniform liner, and 3D SPH with 30 discrete plasma jets. Two dimensional slices of the pressure show that the discrete jet SPH case evolves towards a profile that is almost indistinguishable from the SPH case with a uniform liner, showing that non-uniformities due to discrete jets are smeared out by late stages of the implosion. Liner formation and implosion on vacuum was also shown to be robust to Rayleigh-Taylor instability growth. Interparticle mixing for a liner imploding on vacuum was investigated. The mixing rate was very small until after peak compression for the 30 jet simulation.

I.     Introduction

Imploding "liners" are used for compressing plasma to a high energy density state. In magneto-inertial fusion (MIF)[1,2], both solid[3,4] and plasma liners[5,6] are envisioned to compress plasma to fusion conditions.

The Plasma Liner Experiment (PLX)[7] plans to explore and demonstrate the feasibility of forming spherical plasma liners imploding on vacuum that can generate cm-, μs-, and Mbar-scale plasmas upon stagnation. The plasma liners on PLX will be formed via merging of 30 dense, high Mach number ($M$), pulsed-power driven plasma jets (ion density $n \sim 10^{17}\,\mathrm{cm}^{-3}$, $M \sim 10$–$35$, velocity $V \sim 50$ km/s, jet radius $r_{jet} \sim 5$ cm) in spherically convergent geometry (Fig. 1), with total capacitive stored energy of ~1.5 MJ. In the near term, PLX aims to enable an experimental platform for fundamental studies in high energy density laboratory physics (HEDLP) and laboratory plasma astrophysics, and in the longer term PLX can further explore the potential for imploding plasma liners to be a standoff driver for MIF[6].

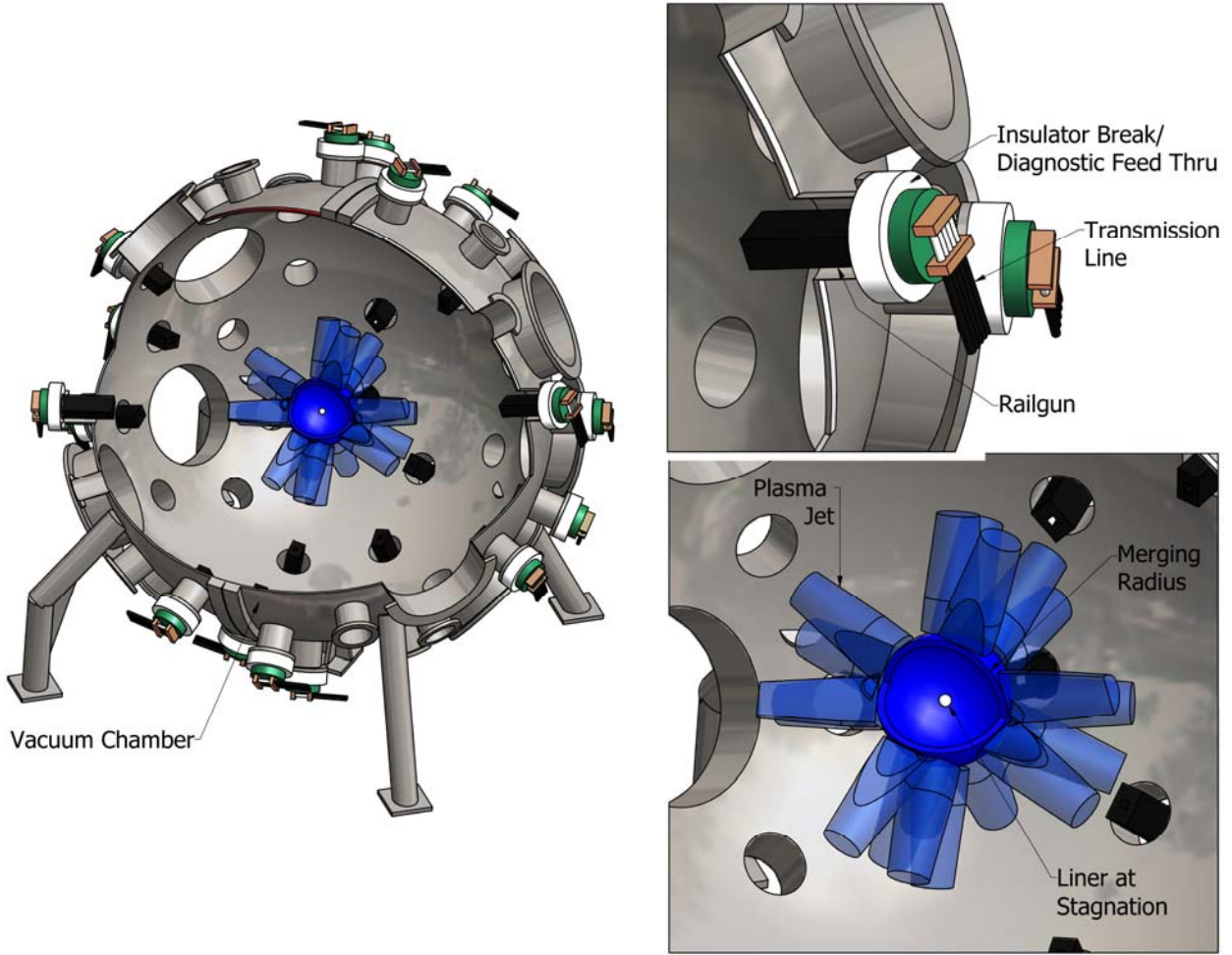

**Figure 1. Illustration of 30 plasma jets produced by plasma guns at the surface of a spherical vacuum chamber forming a spherically imploding plasma liner (figure credit: HyperV Technologies Corp.).**

A key concern for spherically imploding plasma liners to reach high pressure is that perturbations may lead to significant asymmetries in the radial momentum and subsequently to the onset of convergent instabilities. In this paper we utilize SPHC[8], a smooth particle hydrodynamics (SPH)[9] code, to model plasma liner formation, implosion, and expansion. The objectives are to compare results from one-dimensional (1D) and three-dimensional (3D) ideal hydrodynamic simulations using anticipated PLX jet parameters, evaluate conditions for potential onset of Rayleigh-Taylor instabilities, and estimate fluid particle mixing for the 30 jet simulation.

To meet these objectives, we compare the results from a case utilizing 30 cylindrical jets with an equivalent implosion model of a spherically uniform liner simulated using 1D and 3D hydrodynamic codes. In Sec. 2, we present a brief overview of the SPH numerical approach. In order to provide some confidence in the numerical output, 3D verification studies were performed using the Noh shock problem[10,11], and those results are given in Sec. 3. Descriptions of the 30 jet and spherically uniform models are given in Sec. 4. Section 5 summarizes the comparison of pressure and radial history of the 3D liner simulations with results from a set of 1D hydrodynamic simulations reported elsewhere[12]. In Sec. 6 it is shown that the 30 jet implosion tends to evolve towards good spherical symmetry. Sections 7 and 8 analyze possible reasons for this with respect to the onset of Rayleigh-Taylor instabilities and mixing. Conclusions are discussed in Sec. 9.

## II.    Smoothed Particle Hydrodynamics

Modeling of processes in devices for laboratory astrophysics or fusion is challenging because of the wide range of spatial and temporal scales involved under conditions of high energy density. In PLX, for instance, in which the magnetic field is energetically and dynamically negligible, the smallest time scale is the electron/ion scattering time $\tau_{ei} \sim 10^{-12}$ s while the entire experiment occurs over $\sim 10^{-5}$ s. Spatial scales vary from the electron scattering mean free path $\lambda_{ei} \sim 10^{-5}$ m up to the vacuum chamber diameter of 3 m. The interior of the plasma jets is locally collision-dominated, but the edges are sufficiently rarefied that two fluid or kinetic effects may be important. Jet merging may be partially collisionless at the anticipated $10^{23}$ m$^{-3}$ jet densities expected at merging, because the relative merging velocity between jets can give an effective temperature of 50 to 100 eV, with a corresponding $\lambda_{ei,merge}$~1 cm, which is about 10 to 20% of the jet diameter. The thermalized liner will be collision dominated, and may transition from optically thin to optically thick during the implosion. Radiation transport is expected to be an important energy transport mechanism during stagnation, when temperatures will exceed 100 eV. Radiative cooling of the jets may also play a role during liner formation. Thermal transport will be significant, with electron thermal conduction in the hot spot exceeding $10^4$ W/m·K. Since high-Z liners are desired to deliver high momentum flux to magnetized targets in follow-on experiments to PLX, multiple ionization states will be important. Because of the wide range of length and time scales, numerical modeling of the entire experiment is an extremely challenging problem, which cannot be handled by a single numerical code. Our overall strategy in modeling PLX is to combine the particle in cell (PIC), two-fluid, 2D/3D magneto-hydrodynamic (MHD) and 1D/3D ideal and radiation hydrodynamic numerical simulations. This paper will focus on ideal 3D hydrodynamic modeling, with the intent to incorporate tabular equations of state, opacities, and thermal transport in a future work. The purpose of starting with ideal hydrodynamic equations of motion, closed with a constant specific heat ratio and ideal gas law, is to ascertain basic trends in scaling of merging radius, dwell time, and peak pressure with various jet parameters. Validation against experimental data will proceed as data becomes available.

The choice of Smoothed Particle Hydrodynamics (SPH)[13], a free Lagrange method, for 3D plasma liner simulations was made for several reasons. PLX involves the flow evolution of discrete jets each with a fixed amount of material surrounded by vacuum. These jets propagate through a considerable amount of empty space. Eulerian codes have to discretize a large number of nodes, of which most would consist of a vacuum, leading to ineffective use of computational memory and CPU time to match the resolution of a Lagrangian model in which only the jets were discretized. However, Lagrangian codes are susceptible to excessively skewed and entangled meshes in regions of strong shear and turbulent motions, introducing other forms of numerical inaccuracies and/or resulting in premature termination of the computation. Gridless Lagrangian methods retain the accuracy of the Lagrangian step while avoiding the pitfalls of mesh distortion. They accomplish this by discretizing the domain into a discrete set of points as in Lagrange and Eulerian methods, but allowing for the connectivity between neighboring points to change as fluid particles move within the domain during computation[14]. Due to the arbitrary mesh connectivity, meshes can be variably zoned allowing for increased resolution only where it is needed. One of the greatest advantages of SPH, and the main reason for using it in this study, is its relatively short computational time requirements and ability to easily perform 3D calculations. We summarize the basic discretization formulation of SPH here.

SPH was invented by Lucy[15] and Gingold and Monaghan[16], and it has been traditionally used to investigate astrophysical processes, notably the formation of the moon and the fission of stars into binary stars[13]. SPH is a gridless Lagrangian technique[17], in which a differential interpolant of a function can be constructed from its values at the particles by using a differentiable kernel, whereby derivatives are obtained by ordinary differentiation[18]. As in finite element methods, the kernel acts as a differential test or interpolation function. For instance, the integral interpolant of any function is defined by

$$A(\mathbf{r}) = \int A(\mathbf{r}') W(\mathbf{r}-\mathbf{r}', h) d\mathbf{r}' \quad (1)$$

where W is the interpolating kernel, r is the position of the particle, and h is the radius of influence measured from particle a. Numerically, Eq. 1 can be approximated by a summation interpolant

$$A(\mathbf{r}) = \sum_b m_b \frac{A_b}{\rho_b} W(\mathbf{r} - \mathbf{r_b}, h) \qquad (2)$$

where *m* and *ρ* are the mass and density of particle *b*, respectively. Derivatives of *A* are straightforward, and the gradient of *A* is calculated as

$$\nabla A_a(\mathbf{r}) = \sum_b m_b \frac{A_b}{\rho_b} \nabla W(\mathbf{r} - \mathbf{r_b}, h) \qquad (3)$$

### III. Verification of SPHC with the 3D spherical Noh Problem

We verified SPHC against the Noh problem[10,11] to give confidence in the numerical output. Specifically we determined the $L_1$ norm and convergence rates for the pressure, temperature, and density behind the post-shock region of an imploding fluid obeying the ideal gas law. The Noh problem is a self-similar solution in Cartesian, cylindrical, or spherical coordinates in which either a uniform jet impinges on a wall or an infinitely thick, initially uniform, cylindrical or spherical shell collapses on either the axis (cylindrical) or origin (spherical), Fig. 2.

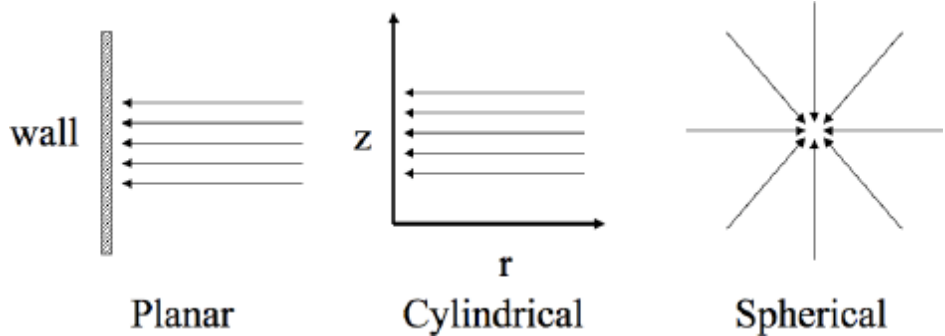

**Figure 2.** Initial velocity vector for 1D planar, cylindrical, and spherical Noh shock problems.

A shock propagates through the incoming flow for all t>0, and the solution gives the shocked density, temperature, and shock speed behind the shock. This case tests the hydrodynamic shock capturing capability of the algorithm, and identifies both dispersion and diffusion errors.

We ran a 3D (symmetric) case to test the numerical output of SPHC. Initial conditions were: $V_0$ = 100 km/s, $T_0$ = 10 K, $\rho_0$ = 1 kg/m³, $MW$ = 1 kg/mol. The analytical solution gives after-shock values of 34.72 eV, 2120.09 kbar and 63.29 kg/m³.

A convergence test was performed for 10000, 20000, 40000 and 80000 particles using the $L_1$ norm to measure the error. Following this, the results are shown in Fig. 3 below. P, T, and ρ, converge at a rate of $n^{0.64}$, $n^{0.46}$, and $n^{0.63}$, respectively, and the errors are all less than 10%. The resolution in the simulations is ~30,000 particles, which is beyond the steep, coarse resolution end of the convergence curve.

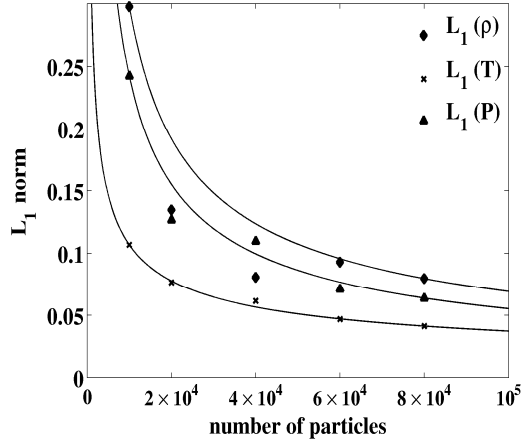

**Figure 3.** $L_1$ norm for 3D Noh test case.

**IV. Description of Uniform and Discrete Jet Models**

Each 3D SPH simulation is either performed as a uniform imploding shell or an equivalent set of discrete jets, with the working fluid being argon obeying a constant gamma law ideal gas. Equivalence between uniform and discrete jet runs means that the total mass (300 mg), kinetic energy (376 kJ), and initial Mach number ($M_0 = 25$) are the same. The uniform case is started at the estimated merging radius ($R_m = 0.241$ m) of the jets with a shell thickness ($\Delta R_0 = 0.255$ m), initial velocity ($v_{M0} = 50$ km/s), mass density ($\rho_{M0} = 6.63 \times 10^{-4}$ kg/m$^3$), and temperature ($T_0 = 1$ eV), Fig. 4a. The velocity vector for all SPHC particles defining a given jet were parallel to the axis of symmetry of that jet, so most particles had a small but finite component of polar and/or azimuthal velocity component. For the discrete jet model, each jet is positioned at the chamber wall radius ($R_C = 1.37$ m), with length ($l_{j0} = \Delta R_0$), Fig. 4b. Thermal expansion will cause the jets to be roughly 1/3 longer at the merge radius than the liner for the PLX conditions of interest, and this effect will be considered later in the paper. The initial temperature and velocity are set to the same values as the uniform case, while the density ($\rho_{j0} = 5 \times 10^{-3}$ kg/m$^3$) is scaled so that the total mass matches that of the uniform liner. To be consistent with the 1D study by Awe et al.[12], the jet radius ($r_{j0}$ = .05 m) is chosen based on the expected size of the jets discussed in that paper. The estimate provided a means for calculating the merging radius for the initialization of the 1D simulations and was selected based on the expected jet radius to be produced by pulsed plasma accelerators[19].

In the 30 jet case, the jets were distributed evenly among the 60 nodes of a truncated icosahedron (soccer ball) with the leading edge of each jet placed at the chamber wall radius, Figs. 4 and 5. There are numerous ways to position the jets on the available nodes, and the criteria for the optimum coverings have yet to be determined. In this example, the buckyball is oriented in spherical coordinates such that the z axis is centered through a pentagon on the top and bottom surfaces of the sphere, Fig. 5, and the nodes fall on constant latitudes of approximately 20.1º, 43.4º, 59.0º, 80.1º, 99.9º, 121.0º, 136.6º, and 159.9º, when the polar angle is measured from the positive z-axis. The corresponding number of nodes at each latitude are 5, 5, 10, 10, 10, 10, 5, and 5, respectively. Referring back to the simulation represented in Fig. 4, the 30 jets are arranged with 5 jets at 20.1º, 10 at 59.0º, 10 at 121.0º, and 5 at 159.9º.

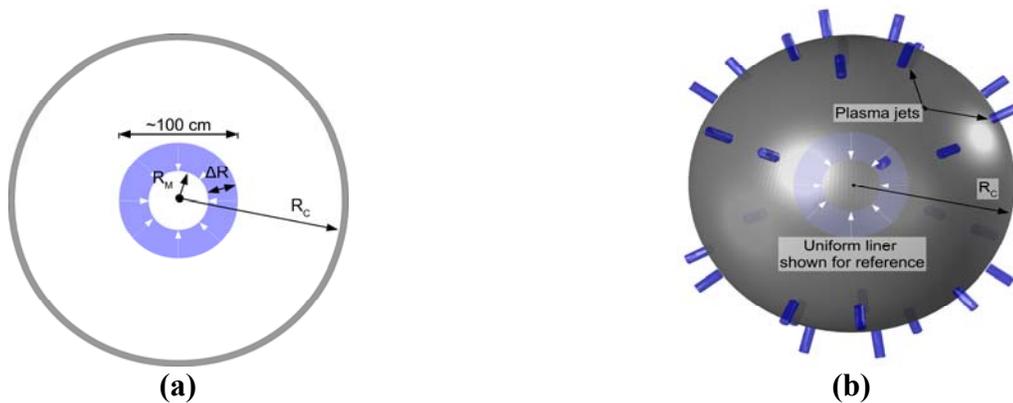

**Figure 4. Uniform (a) and discrete jet (b) initial setup for liner implosion modeling. The uniform liner is shown in (b) for reference only.**

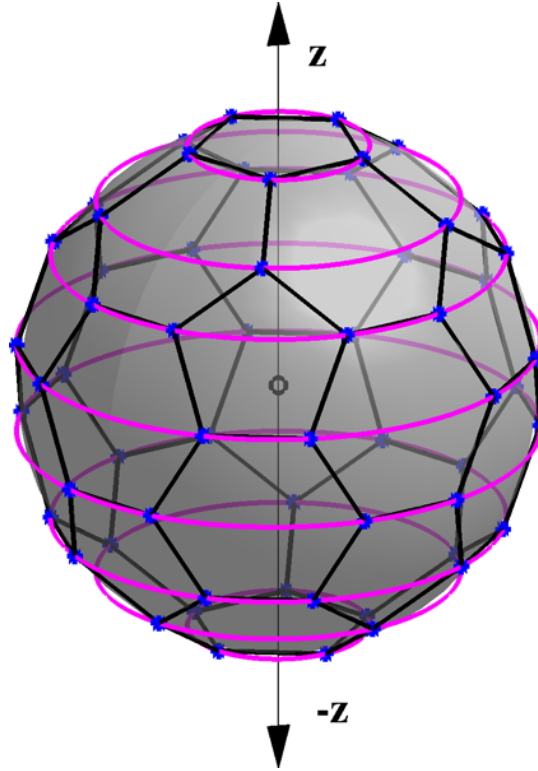

**Figure 5.** Nodes on a truncated icosahedron (buckyball), showing lines of constant latitude.

## V. Comparison of 1D and 3D Simulations

A series of 1D runs[12] was conducted using the two radiation hydrodynamics codes RAVEN[20] and HELIOS[21]. The initial conditions for the uniform and 30 jet cases in the present 3D study, as described in Sec. , are equivalent to those of the 1D run 6 of Table 2 of Ref. 12, and are based on the anticipated experimental conditions of PLX.

The pressure and radius are measured at a point inside the liner at a depth of 2 particle spacings (2h), averaged over the 4π solid angle, which is equivalent to the results from the 1D models, which corresponds to values in the cells of initially 1 mm depth. To show this is an equivalent measurement, for a 1D simulation like RAVEN, one has the luxury of selecting a specific cell and tracking it in time. In 3D SPH, the particles move about in accordance to conservation laws in a Lagrangian frame. Since particles can interpenetrate and mix, it would not be equivalent to the 1D simulations to pick a particular set of particles and plot the properties. Further, the choice of calculating the pressure at a depth of 2h from the inside of the liner averaged over the solid angle allows a very good estimate of the peak pressure history while avoiding edge effects due to SPH interpolation.

We compare the time history of the pressure at the inner liner among the equivalent 1D and 3D runs, Fig. 6. The radius vs. time is shown for reference. The initial pressures vary considerably, but as the liner converges, the pressure curves coalesce around 4.5 μs. In Ref. 22, we estimated the dwell time $\tau_{dwell}$ for a target confined by an imploding liner to be $\sim l_{j0}/v_{M0}$ using a self-similar converging shock model. In all cases in the present study (which does not have a target), the pressure peaks generating a hotspot which diminishes over a period of 0.5 to 1 μs, roughly 10% to 20% of $l_{j0}/v_{M0}$, and the percentage is similar to the bounce back time in laser driven inertial confinement fusion (ICF). The post shocked region is maintained at a pressure of roughly 10% of the peak, for a duration on the order of $\tau_{dwell}$.

Note that during the stagnation period, the pressure of both 3D cases are in between those of the 1D simulations. Further, from 6.5 to 9 μs, the pressures of the 3D uniform and 30 jet cases are almost identical. Finally, note that the 30 jet case appears to have a slightly higher peak pressure, and an overall longer confinement time compared to the uniform 3D liner simulation. The longer confinement time is explained in the next section.

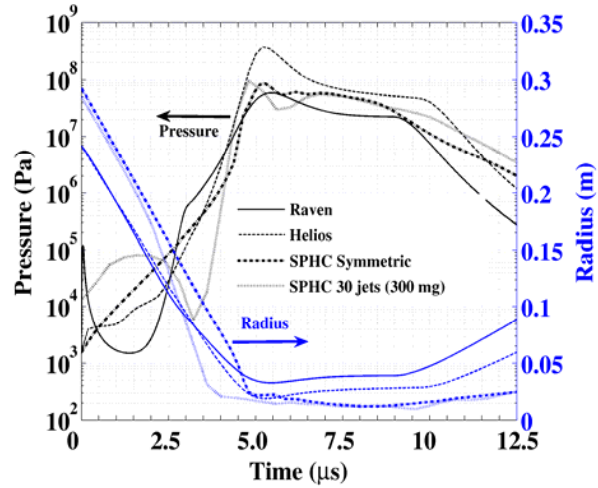

**Figure 6.** *P(t)* and *r(t)* **curves. The 1D ideal hydrodynamic RAVEN and HELIOS data are results from case 6 of Table 2 Ref. 12.**

## VI.    Comparison of Uniform with Discrete Jets Cases

To evaluate the importance of 3D effects associated with discrete jets, 3D SPH simulations with an initially uniform liner and 30 discrete jets are compared.  The striking similarities in the pressure history between the two cases suggest that the physics of liner formation by discrete jets does not introduce instabilities or other processes that would lead to significant departures from the uniform case when the liner implodes on vacuum.  This can be observed comparing contour plots of pressure at equivalent times for both simulations, Figs. 7-10.  Locally, the pressure inside the jets is considerably higher than that of the uniform liner since the total mass and thermal energy is the same, while the total volume of the jets is smaller.  Also, pressure spikes are observed between pairs of jets during the beginning stages of jet merging.  These spikes are caused by the oblique shock layer which forms at the slip surface between merging sets, creating layers with pressure ~30 times that of the undisturbed jets.  This layer of increased pressure redirects material in the shock layer in a direction primarily tangent to the slip surface, somewhat like the motion of a jet impinging on an imaginary wall at the merging half angle. A portion of this material expands in the radial direction towards the origin, creating so-called precursor jets which partially fill the cavity in advance of the incoming liner. As will be shown, the relieving effect will help reduce the amplitudes of the relative pressure perturbations to levels small compared to the stagnation region.

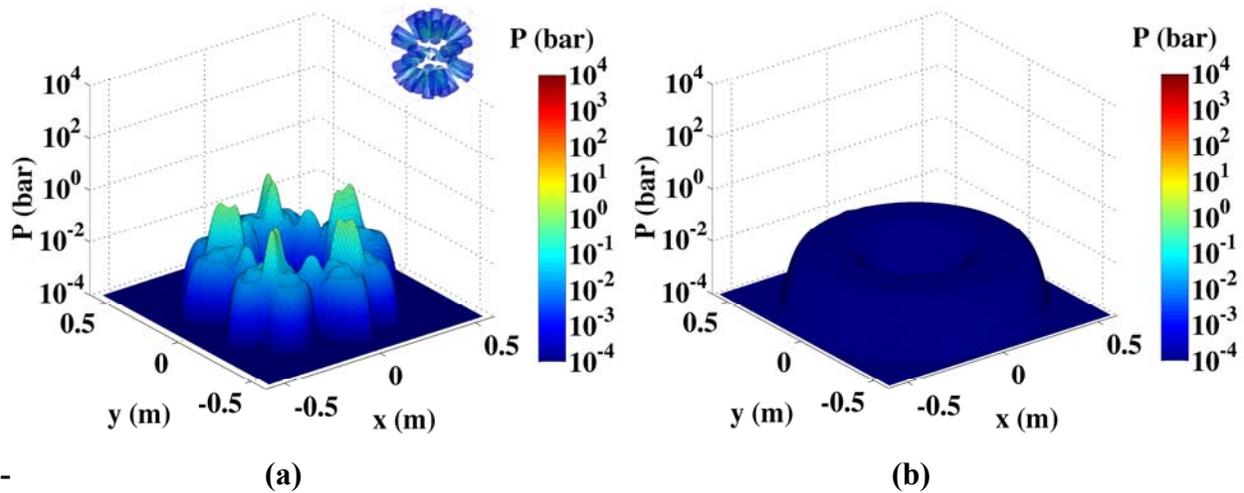

(a)            (b)

**Figure 7. Pressure contours in x-y plane for t = 0 μs for (a) discrete jets (z plane at 0.2 m) and (b) uniform liner. Jet positions shown in (a) for reference.**

By 3.6 μs, the relative peak to value amplitude $\delta p / p$ of the inner liner is dropped to about 4.0, Fig. 8. Peak pressures are also much closer in magnitude between the discrete and uniform cases. The uniform liner is noticeably thicker, because thermal expansion can only occur in the radial direction for uniform shells.

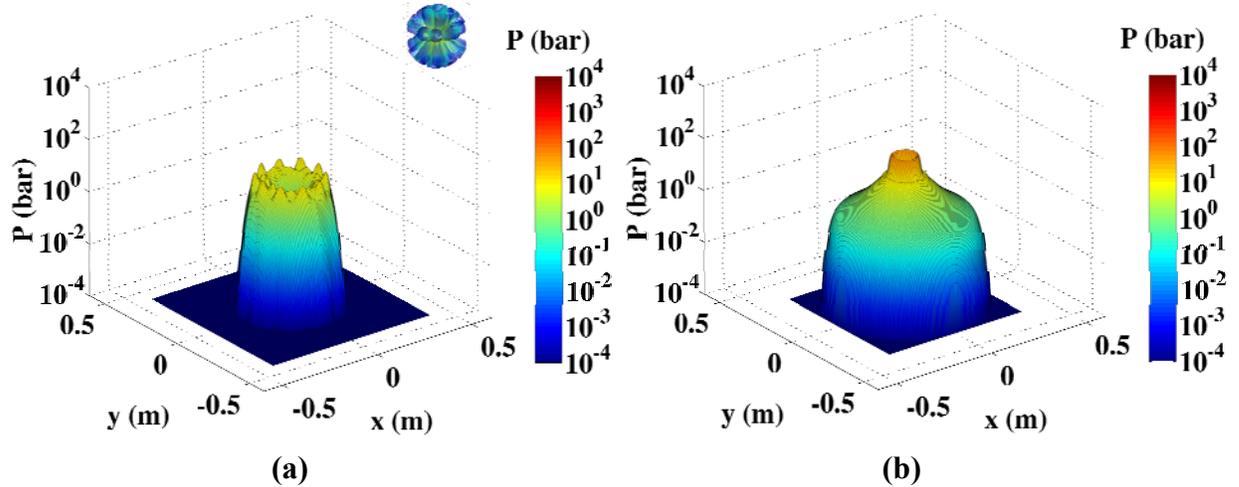

(a)                                (b)

**Figure 8. Pressure contours in x-y plane for t = 3.6 μs for (a) discrete jets (z plane at 0.0 m) and (b) uniform liner. Jet positions shown in (a) for reference.**

Peak pressure occurs in the 30 jet case at 4.8 μs. The pressure profile in the z=0 plane begins to look very similar between the uniform and 30 jet cases, Fig. 9. Rising sharply from the outer liner edge, the pressure gradient becomes less steep for a few cm before again rising sharply towards the stagnated region at the center at the location of the hotspot. This feature is clearly visible in both cases. Remnants from the oblique shock layers are still visible in the 30 jet case, but the perturbation pressures are ~2% of the hotspot pressure. By 8.0 μs, the qualitative pressure profiles are difficult to distinguish, Fig. 10. Note that the uniform case expands more rapidly during the post stagnation phase. At the start of the implosion compressional heating is ubiquitous in the uniform liner, whereas in the discrete jets compressional heating is localized to the region from the leading edge to the merging radius. Thus, the mean temperature of the uniform liner is higher at stagnation and consequently the rarefaction waves that cause expansion travel faster. Finally, we note this result may not be true for all discrete jet implosions under all conditions.

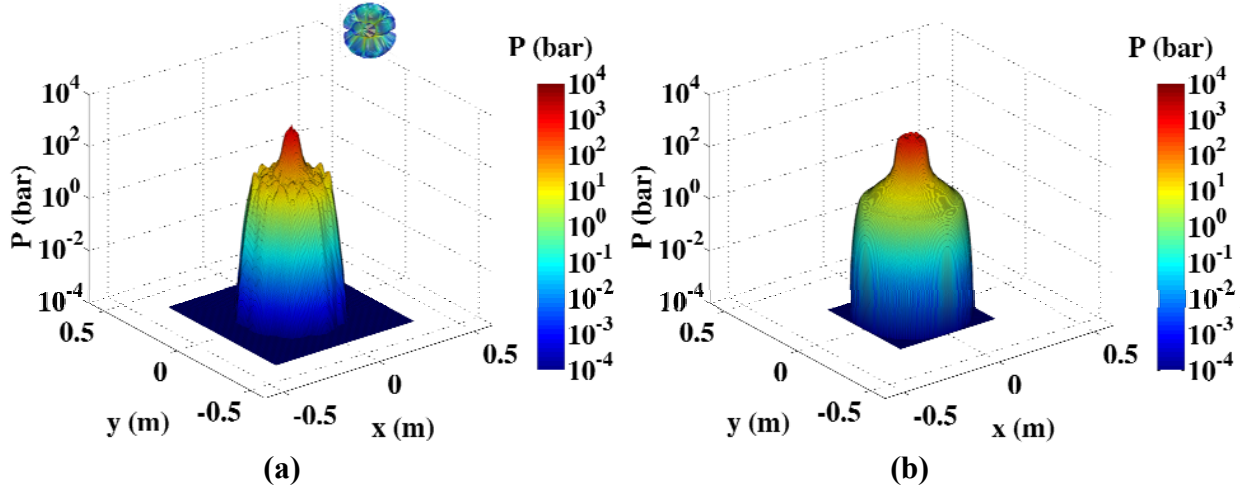

(a)                                    (b)

**Figure 9.** Pressure contours in x-y plane for t = 4.8 μs for (a) discrete jets (z plane at 0.0 m) and (b) uniform liner. Jet positions shown in (a) for reference.

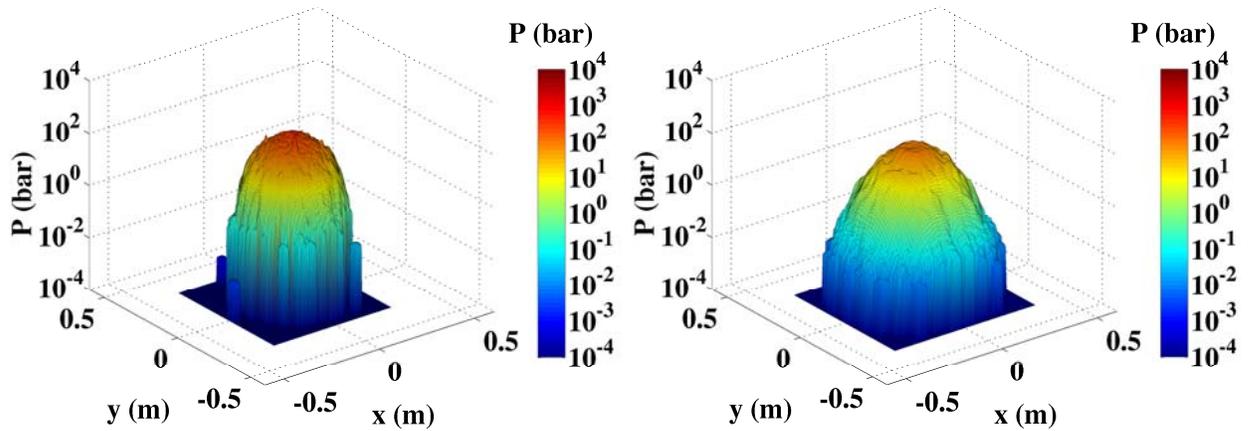

**Figure 10.** Pressure contours in x-y plane for t = 8.0 μs for (a) discrete jets (z plane at 0.0 m) and (b) uniform liner.

## VII. Potential Development of Rayleigh-Taylor Instabilities

The tendency towards formation of a uniform liner appears encouraging. However, it would be constructive to consider conditions at which the jets and the plasma liners may become unstable. Such instabilities, being hydrodynamic in nature, may develop during the implosion process and may significantly amplify the imperfections. Such imperfections are always present in realistic 3D systems, and thus may influence liner formation and liner-target interaction. Extensive studies of plasma fusion, conducted in the several past decades, suggest that it is essential to check if in the PLX system the conditions are met for the development of instabilities of the Rayleigh-Taylor, Richtmyer-Meshkov and Bell-Plesset types. Characteristic features of these instabilities are that they grow faster for smaller wavelength of the imperfections, and at advanced stages of their evolution small scales are strongly coupled to large scales.

Rayleigh-Taylor instability (RTI) develops when materials of different densities are accelerated against the density gradient. Extensive interfacial material mixing is ensured with time. The conditions for development of RTI are commonly implemented in the inequality

$$\nabla P \cdot \nabla \rho < 0 \qquad (4)$$

which was first suggested by Chandrasekhar[23]. Here $P$ and $\rho$ are the fluid pressure and density. For compressible materials, especially in high energy density plasmas, the condition (4) should also be augmented with other considerations[24-26]. Note also that a finite equilibrium pressure should be maintained by the material or by the magnetic field in order for RTI to develop. We refer the reader to a review article[27] and to recent papers[28-30] for an overview of the state of the art in studies of RTI from atomistic to macroscopic scales.

Using our numerical simulations results, we present the value of $\nabla P \cdot \nabla \rho$ for the 30 jet case at times of 0, 3.6, 4.8, and 8.0 μs, see Fig. 11. It should be noted that these plots consist of the slices in the xy plane at a constant value of z, with both the elevation and color corresponding to the value of $\nabla P \cdot \nabla \rho$. The jets are observed to generate regions of instability prone areas during merging. However, by 3.6 μs, when the inner layer of the liner is nearing the center, only the rarefied material at the origin is unstable. Localized regions near the center continue to develop as the implosion and expansion progresses, but these regions do not appear to be disruptive to the overall dynamics. The high degree of isotropy of the dynamics of merging jets is likely due to the fact that the liner effectively implodes on a vacuum as the material at the origin is rarefied and magnetic field is absent. For a finite material density at the origin, and/or in the presence of magnetic field, the anisotropy of the jet merging process can be more pronounced.

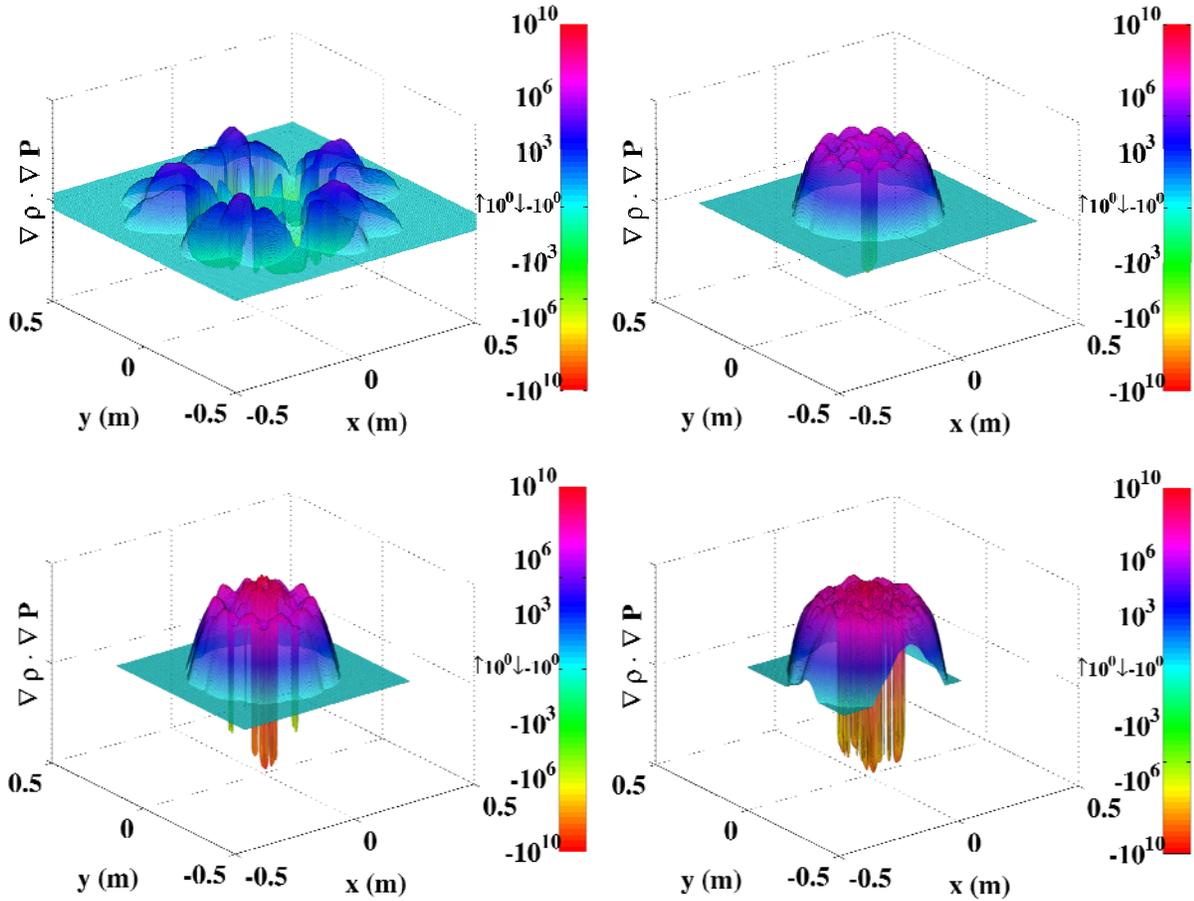

**Figure 11. Evaluation of the Rayleigh-Taylor hydrodynamic condition for stability for (a) 0, (b) 3.6, (c) 4.8, and (d) 8.0 μs.**

We also anticipate that for MIF, target compression by the liner can be RT unstable at the liner/target interface when the liner begins to decelerate, similar to the interface between the hotspot and cold fuel layer in ICF capsules. Criteria for tolerable thresholds, such as the effect of instability modes and relative perturbation amplitude on yield, will be considered in our future work.

We further recognize that the shocks, which may develop at the regions of the jet merging and stagnation, may lead to the development of the Richtmeyer-Meshkov instabilities (RMI), whereas imperfect spherical convergence may lead to Bell-Plessett effects[27-30]. In addition, a sudden deceleration of the imploding liner by a low density magnetized target may be susceptible to the growth of the magnetic RTI (see, *e.g.* Ref. 31). Detailed consideration of these processes is beyond the scope of our present study, and will be a subject of future investigations.

SPHC is an adequate numerical tool to accurate quantification of the heterogeneous, anisotropic and statistically unsteady dynamics under extreme conditions of high velocity, high Mach number and high density contrasts, which are relevant to the PLX. Figure 12 shows the results of one of the test SPHC runs of the nonlinear RMI in planar geometry for $M$=10 and the ratio of the densities of the light and heavy fluid 1:39, taken from Ref. 32. As seen in Fig. 12, at such extreme conditions, the SPHC simulations adequately describes the large-scale dynamics[28-30] and help to identify some new features of RMI evolution, such as the appearance of local heterogeneous micro-structures and a complicated character of the scale-coupling[32].

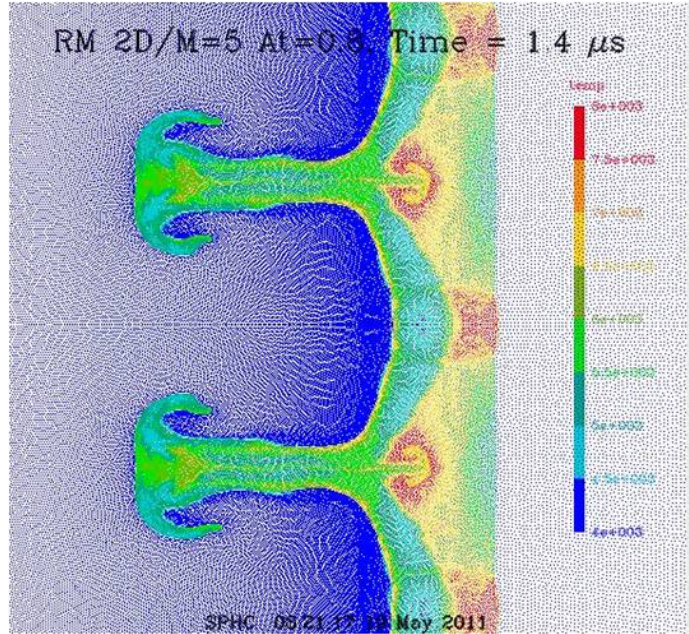

**Figure 12. Richtmeyer-Meshkov instability in planar geometry by SPH for strong shock with Mach number 10 and for gases with highly contrasting density ratio 1:39, from Ref. [32]. Temperature plot in range between 4000K and 8000K, clearly shows local heterogeneous micro-structures and a complicated character of the coupling between the large and small scales.**

Finally, note that the results of recent theoretical analysis[28-30] indicate that inhomogeneous and anisotropic RT mixing exhibits more order compared to canonical turbulent processes. In particular, an accelerated mixing flow has a tendency to become laminar again and has stronger level of correlations, weaker contribution of fluctuations, steeper spectra, and stronger dependency on the initial conditions[28-30]. This opens new opportunities for control and mitigation of RTI and mixing in high energy density plasmas and in PLX-relevant conditions.

**VIII. SPH particle mixing during implosion**

Mixing mechanisms that could possibly occur in plasma jet driven magneto-inertial fusion (PJMIF)[5,6] during the plasma liner collapse could be detrimental to the achievable peak pressure. Increased turbulent mixing are likely to enhance heat transfer via convection therefore lowering peak temperature and pressure and reducing the effective radius of the hotspot. One of the benefits of the SPH Lagrangian particle approach is that each individual particle can be traced through both space and time, therefore providing a direct insight to the particle's history. This has been exploited for evaluating the possibility of mixing in PLX-relevant cases of plasma liner implosion on vacuum.

To better understand the material mixing analysis, the post processing procedure must first be described. Most important is the understanding of how SPHC creates initial plasma jets. Each jet was approximated by a cylinder with dimensions of radius $r = 5$ cm and length $L = 25.5$ cm. A total of 30 symmetrically distributed jets across a sphere were used and total number of the particles in the simulations was 28800, or 960 particles per jet. Each jet is constructed from a set of chords, where a chord is a collinear set of particles at a fixed distance from the jet's axis of symmetry, Fig. 13. For this particular resolution, there were 48 chords per jet, with 20 particles per chord.

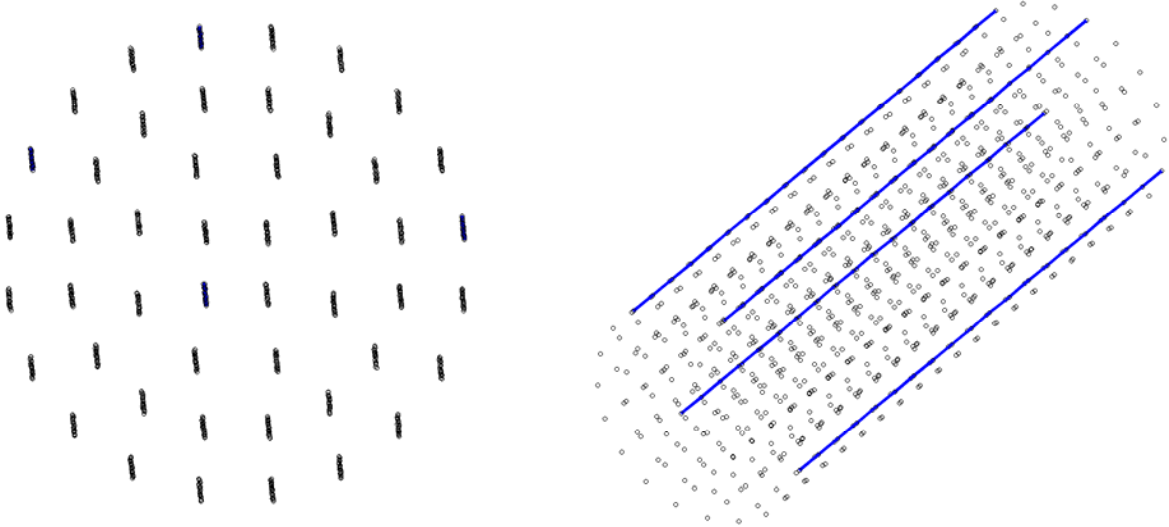

**Figure 13. Front and size facing views of a jet showing four chords (top, northwest, center, and right) that have been analyzed.**

Mixing was determined by a comparison of the radial positions of a particle along a particular chord, which for illustrative purposes we number from 1 to 20, with 1 being the innermost particle. A mixing event is defined as when the $n^{th}$ particle overtakes the $m^{th}$ particle along a single chord, with particle $n$ having initially a larger radial position vector than the $m^{th}$ particle from a previous timestep. To make the mixing calculations tractable, we evaluated mixing events for four chords (top, northwest, right, and center, Fig. 13) each in six jets. The chord positions and jets were selected to probe mixing throughout the domain of the imploding liner for the duration of the simulation, in both regions of relatively high and low gradients.

It is important to state that only mixing in the radial direction was analyzed (with the center of the chamber being the origin of the coordinate system). This is due to assumption that the convection caused by the transverse movement of the particles along the same-radius-sphere would be negligible in comparison to the convection caused by radial mixing, which would cause hot, inner portions of the plasma to move outward. Other metrics to describe mixing may be explored in the future.

We plotted the mixing events recorded by the event location in radius and time, starting with the time at which material begins to fill the center of the cavity inside the liner $t_{collapse}$, Fig. 14. The 'mixing cloud' represents the location in radius and time of all mixing events. Included in this chart is a flow diagram (radius vs. time) of the innermost and outermost particles for some select chords. The purpose of this is to show approximately the trajectory of the liner boundaries as a function of time with respect to the mixing cloud. Mixing begins just before the time of peak pressure, $t_{peak}$. The mixing rate is relatively small for 1 μs beyond the peak pressure, then increases substantially until the stagnation time $t_{stag}$, when the outer portion of the liner has been stopped by the outgoing stagnation shock. At this point, the liner is completely thermalized and begins to expand. The mixing cloud density appears to decrease with time throughout the remainder of the simulation.

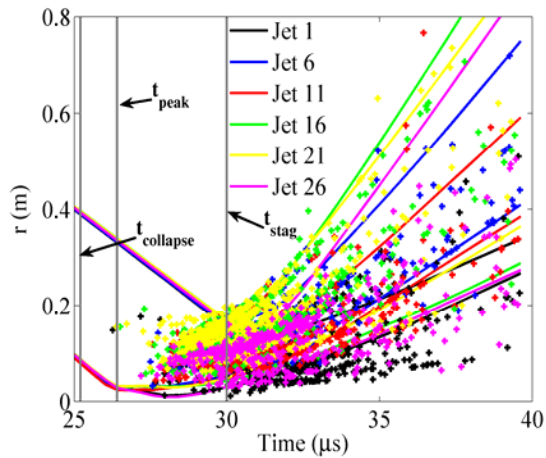

**Figure 14. Flow diagram and mixing cloud for 4 chords on each of 6 jets.**

Encouraged by the relatively late onset of particle mixing, the mixing frequency per particle was plotted to try and gain insights into the development of mixing during the implosion and expansion phases of the simulation. Mixing frequency per particle is shown in Fig. 14 as a function of the dimensionless time parameter

$$T \equiv \frac{t - t_{collapse}}{t_{stag} - t_{collapse}} \tag{5}$$

The probability of a mixing event to occur per SPH particle was found to be described accurately by this parameter with the piecewise function

$$f = \begin{cases} T^6 e^{-\frac{3\pi T^6}{2}}, & \text{for } t \leq t_{stag} \\ \frac{2}{T^3}, & \text{for } t > t_{stag} \end{cases} \tag{6}$$

This function is plotted in Fig. 15 and is in very good agreement with the particle mixing rate determined from the simulation. At the moment of the cavity collapse of the liner, the mixing rate is very low and remains relatively small until roughly half the liner has reached stagnation. Beyond this time, the frequency increases significantly. By stagnation, each SPH particle has on average undergone approximately 1.7 mixing events. For $t \leq t_{stag}$, the mixing distribution function is similar to the Maxwellian speed distribution for $T^3$. Interestingly, the power law behavior for $t > t_{stag}$ is still a function of $T^3$.

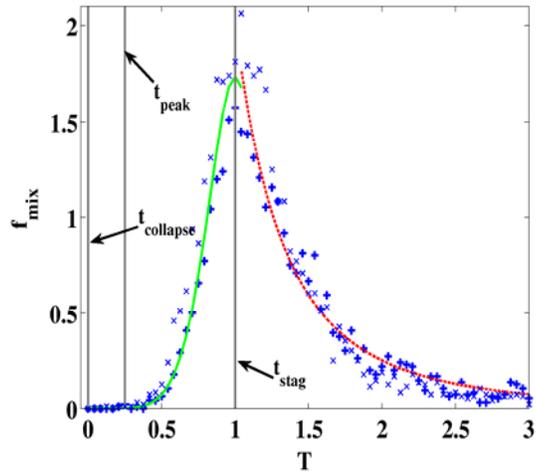

**Figure 15. Mixing frequency as a function of dimensionless time (+ *artificial viscosity*, × *Balsara term included from Ref. 34*).**

Some mention should be made about the possibility of artificial mixing in SPH models. As SPH is a particle-based method, artificial viscosity is required to prevent particles from inter-penetrating and passing by each other for unphysical reasons. This scenario often occurs in regions of large pressure gradients such as in shocks. In those regions, particles have a tendency to pass by each other or merge, which relaxes the gradient or causes code instability. Such behavior is unphysical and artificial viscosity terms are needed[33]. Most artificial viscosity models influence equally both shear and normal stresses. As shear stress is the main driver of the Kelvin-Helmholtz instability, which increases mixing, additional analysis with the alternative Balsara artificial viscosity[34] was done in order to ensure the choice of artificial viscosity does not introduce artifacts in the particle mixing results. Balsara came up with a method that decreases the influence of shear stress on particle acceleration. After comparing the two simulations (one with Balsara term on and one without it), no significant deviations in mixing frequency was noticed, but both sets of data are included in Fig. 15. Therefore it can be concluded that for purposes of simulating the mixing in plasma liner implosions with given resolution, regular artificial viscosity terms should be satisfactory.

## IX. Conclusions

Three-dimensional hydrodynamic simulations have been performed using smoothed particle hydrodynamics to study the effects of discrete jets on the processes of plasma liner formation, implosion, and expansion. Comparisons among two 1D hydrodynamic simulations (RAVEN and HELIOS), a 3D uniform simulation in SPH, and a 30 jet simulation show that in all cases the pressure peaks during liner collapse and decreases rapidly for a period of 0.5 to 1 µs. Subsequently the pressure remains almost flat until the outer edge of the liner stagnates. From 6.5 to 9 µs, the pressures of the 3D uniform and 30 jet cases are almost identical. The 1D and 3D simulations were in good agreement and suggest that formation of a liner by discrete jets does not necessarily compromise performance. Comparisons of 2D slices of the pressure profile showed that the plasma liner formed by discrete jets evolves in time towards the case with an initially uniform plasma liner. An examination of the criteria for onset of Rayleigh-Taylor instabilities revealed short-lived instability prone regions at jet merging, and a more pronounced region at the origin during liner collapse. Development of instabilities during target compression by the liner is thus anticipated, but liner formation and implosion appears to be robust to instabilities. Mixing was determined to be negligible until 1 µs after peak pressure. Prior to complete stagnation of the liner, the mixing frequency per particle could be described by a Maxwellian-like expression as a function of a dimensionless time parameter, and afterward by a power law. Overall it appears that liner formation and implosion is a stable process. Further work needs to be done to understand the potential instabilities which could develop for the case of liner compression of a target for MIF.  Effects from ionization, thermal conduction, radiative transfer, and the presence of magnetic fields also need to be considered.  Finally, the mixing which develops may convect heat away from the hot spot and needs to be studied.

**Acknowledgments**

This work was supported in part by the Office of Fusion Energy Sciences of the U.S. Dept. of Energy under Grants DE-SC0003560 and DE-FG02-05ER54810.